\documentclass[journal]{IEEEtran}
\usepackage{mathrsfs}
\usepackage{pifont}
\usepackage{bbding}
\usepackage{amsmath}
\usepackage{tipa}
\usepackage{algorithm}
\usepackage{algorithmic}
\usepackage{amsmath,epsfig}
\usepackage{amssymb}
\usepackage{amsfonts}
\usepackage{epic}
\usepackage{graphicx}
\usepackage{curves}
\begin{document}
\newtheorem{theorem}{Theorem}
\newtheorem{proposition}{Proposition}
\newtheorem{definition}{Definition}
\newtheorem{lemma}{Lemma}
\newtheorem{corollary}{Corollary}
\newtheorem{remark}{Remark}
\newtheorem{construction}{Construction}

\newcommand{\supp}{\mathop{\rm supp}}
\newcommand{\sinc}{\mathop{\rm sinc}}
\newcommand{\spann}{\mathop{\rm span}}
\newcommand{\essinf}{\mathop{\rm ess\,inf}}
\newcommand{\esssup}{\mathop{\rm ess\,sup}}
\newcommand{\Lip}{\rm Lip}
\newcommand{\sign}{\mathop{\rm sign}}
\newcommand{\osc}{\mathop{\rm osc}}
\newcommand{\R}{{\mathbb{R}}}
\newcommand{\Z}{{\mathbb{Z}}}
\newcommand{\C}{{\mathbb{C}}}
%

\title{Genetic Algorithm Based Nearly Optimal Peak Reduction Tone Set Selection for Adaptive
Amplitude Clipping PAPR Reduction}
\author{{Yajun~Wang, Wen~Chen,~\IEEEmembership{Senior Member,~IEEE}, and
Chintha Tellambura,~\IEEEmembership{Fellow,~IEEE}}
\thanks{Yajun~Wang and Wen Chen are with Department of Electronic Engineering,
Shanghai Jiao Tong University, Shanghai, 200240, P. R. China. Wen
Chen is also with SKL for ISN, Xidian University, P. R. China,
e-mail: \{wangyj1859;wenchen\}@sjtu.edu.cn.}
\thanks{Chintha Tellambura is with Department of Electrical and Computer
Engineering, University of Alberta, Edmonton, Canada,  T6G 2V4.
e-mail: chintha@ece.ualberta.ca.}
\thanks{This work is supported by NSFC \#60972031, by national 973 project
\#2012CB316106 and \#2009CB824900, by NSFC \#61161130529, by
national key laboratory project \#ISN11-01.}}


%

\markboth{IEEE Transactions on Broadcasting, vol. X, no. XX, 2012
}{Shell \MakeLowercase{\textit{et al.}}: Bare Demo of IEEEtran.cls
for Journals }
\maketitle

\begin{abstract}
In tone reservation (TR) based OFDM systems, the peak to average
power ratio (PAPR) reduction performance mainly depends on the
selection of the  peak reduction tone (PRT) set and the optimal target
clipping level. Finding the optimal PRT set
requires an exhaustive search of all combinations of  possible PRT
sets, which is a nondeterministic polynomial-time (NP-hard) problem, and this search
is infeasible for  the number of tones used in practical systems.  The
existing selection methods, such as the consecutive PRT set, equally
spaced PRT set and random PRT set, perform poorly compared to  the optimal PRT
set or incur high computational complexity. In this paper, an
efficient scheme  based on genetic algorithm (GA) with lower
computational complexity is proposed for  searching a   nearly optimal PRT
set. While  TR-based  clipping  is simple and
attractive for practical implementation,  determining  the
optimal target clipping level is difficult. To overcome this
problem, we propose an adaptive clipping control algorithm.
Simulation results show that our proposed  algorithms  efficiently obtain a
nearly optimal PRT set and good PAPR reductions.
\end{abstract}



%
\IEEEpeerreviewmaketitle

\section{Introduction}\label{sec:1}
Orthogonal frequency division multiplexing (OFDM) is widely used  in high-speed wireless
communication systems because of its inherent robustness against multipath fading  and resistance to narrowband interference~\cite{IEEEconf:1}. However,   OFDM suffers from  the
high peak to average power ratio (PAPR) of the transmitted signal. This issue  can cause serious problems including a severe power penalty
 at the transmitter.
Conventional solutions to reduce the  PAPR are to use a linear amplifier
or to back-off the operating point of a nonlinear amplifier. But
both these  solutions result in a significant loss of power efficiency.
Many methods have thus  been proposed to reduce the PAPR by modifying
the signal itself.
The simplest one  is clipping the OFDM signal below a PAPR threshold
level~\cite{IEEEconf:2,IEEEconf:3}, but it degrades the
bit-error-rate (BER) of the system and results in out-of-band noise
and in-band distortion. Coding~\cite{IEEEconf:4} is another
technique. Although it can offer the best PAPR reductions, the
associated complexity and data rate reduction limit its application.
Selected mapping (SLM) technique~\cite{IEEEconf:5} and the partial
transmit sequence (PTS)~\cite{IEEEconf:6} are based on multiple
signal representation method. These methods
\cite{IEEEconf:7,IEEEconf:9,IEEEconf:11,IEEEconf:12} improve the
PAPR statistics  of the OFDM signals, but  side information  may be
transmitted from the transmitter to the receiver, which results in a
loss of data throughput.


By modifying the modulation constellation, the active set extension
(ASE) method \cite{IEEEconf:13}, the adaptive  active set
extension~\cite{IEEEconf:15} and the constellation extension
method~\cite{IEEEconf:14}  reduce PAPR, but
these algorithms require increased   power and
computational complexity at the transmitter.

The tone reservation (TR)
technique~\cite{IEEEconf:16,IEEEconf:17,IEEEconf:18}  proposed by
Tellado is a distortionless method based on using a small subset of
subcarriers, called peak reduction tones~(PRTs), to generate a
peak-canceling signal for  PAPR reduction. The method is simple,
efficient and does not require transmission of side information. The
tone reservation technique can be divided into two classes: 1)
TR-gradient-based technique; 2) TR-clipping-based technique, which
is our major focus  in this paper. The PAPR reduction performance of
the TR-clipping-based technique mainly depends on the selection of
peak reduction tone (PRT) set and the optimal target clipping level.
The optimal PRT set will result in the best PAPR reduction. However,
finding the optimal PRT set is a nondeterministic polynomial-time
(NP-hard) problem and cannot be solved for the number of tones
envisaged in practical systems.  So suboptimal solutions are
typically preferable, such as the consecutive PRT set, equally
spaced PRT set and random PRT set. Although the performance of
random PRT set outperforms those of consecutive PRT set and equally
spaced PRT set, it requires enough larger PRT set  sampling to
obtain better PAPR reduction. The cross entropy (CE)-PRT algorithm
in~\cite{IEEEconf:20,IEEEconf:33} obtains better secondary peak, but it requires
larger population or sampling. On the other hand, determining  the
optimal target clipping level, which directly affects  the PAPR
reduction of the TR-clipping-based technique, is also difficult,
because many factors, such as the number of OFDM subcarriers, the
location of PRT set and the modulation scheme significantly
influence  the selection of the optimal target clipping level.

In this paper, we first propose a new suboptimal PRT set selection
scheme based on the genetic  algorithm (GA), which can efficiently
solve the NP-hard problem. An adaptive amplitude clipping (AAC-TR)
algorithm is also developed to obtain good PAPR reduction
performance regardless of the initial target clipping level.
Simulation results show that the GA optimization scheme  achieves a
nearly optimal PRT set and  requires far less computational
complexity than the random PRT set method. The proposed AAC-TR
algorithm also achieves  good PAPR reduction performance.

This paper is organized as follows. In Section~II, the system model
based on the TR method is  introduced and the principles of TR
techniques are described. The GA algorithm for  the nearly
optimal PRT set is proposed in Section~III. In Section~IV, the
adaptive amplitude clipping (AAC-TR) algorithm  is developed. The performances of GA
algorithm, AAC-TR and other algorithms for PRT selection and PAPR
reduction are evaluated by computer simulations in Section~V.
Conclusions are made in Section~VI.

In this paper, $\parallel\cdot\parallel$ denotes the mean square
norm of a vector. $\parallel\cdot\parallel_\infty$ denotes the
$l^\infty$ norm of a vector. $E[\cdot]$ denotes the expectation of a
random variable. $\bar x$ denote the complex conjugate of a complex
number $x$. $(\cdot)^T$ denotes the transpose of a matrix.
$(\cdot)^H$ denotes the conjugate transpose of a matrix.

\section{OFDM Systems And Tone Reservation Technique}\label{sec:2}
This section will describe  the orthogonal frequency division multiplexing (OFDM) signal,
the peak-to-average power ratio (PAPR) and the tone reservation technique.

\subsection{OFDM Systesms and PAPR}
An OFDM signal is  the sum of $N$ independent, modulated  tones
(subcarriers) of equal bandwidth with frequency separation $1/T$,
where $T$ is the time  duration  of the OFDM symbol. For a
complex-valued phase-shift keying (PSK) or quadrature amplitude
modulation (QAM) input OFDM block
$\textbf{X}=[X_0,X_1,\ldots,X_{N-1}]^T$ of length $N$, the  inverse
discrete Fourier transform (IDFT)  generates the ready-to-transmit
OFDM signal. The discrete-time OFDM signal is expressed as
\begin{equation}\label{eq1}
x_{n}=\frac{1}{\sqrt{N}}\sum_{k=0}^{N-1}X_k\cdot e^{\frac{j2\pi
nk}{N}},\quad n=0,1,\cdots,N-1,
\end{equation}
which can also be written in matrix form
$\textbf{x}=[x_0,\dots,x_{N-1}]^T=\textbf{Q}\textbf{X}$, where
$\textbf{Q}$ is the IDFT matrix with the $(n,k)$-th entry
$q_{n,k}=\frac{1}{\sqrt{N}} e^{\frac{j2\pi nk}{N}}$.

The PAPR of $\textbf{x}$ is defined as the ratio of the maximal
instantaneous power to the average power; that is
\begin{equation}\label{eq2}
PAPR(\textbf{x})=\frac{\underset{0\leq n< N}{\max}|x_{n}|^2}{E[|
x_n|^2]}.
\end{equation}

The complementary cumulative distribution function (CCDF) is one of
the most frequently used performance measures for PAPR reduction,
representing the probability that the PAPR of an OFDM symbol exceeds
the given threshold $PAPR_{0}$, which is denoted as
\begin{equation}\label{eq3}
 CCDF=Pr(PAPR>PAPR_{0}).
\end{equation}

\subsection{ Tone  Reservation Technique}
In the TR-based OFDM  scheme,  peak reduction
tones (PRT) are reserved to generate PAPR reduction signals. These
reserved tones do not carry any data information, and they are only
used for reducing PAPR.  Specifically, the peak-canceling signal
$\textbf{c}=[c_{0},c_{1},\ldots,c_{N-1}]^T$ generated by reserved
PRT  is added to the original time domain signal
$\textbf{x}=[x_{0},x_{1},\ldots,x_{N-1}]^T$ to reduce its PAPR. The
PAPR reduced signal can be expressed as
\begin{equation}\label{eq4}
 \textbf{a}=\textbf{x}+\textbf{c}=\textbf{Q}(\textbf{X}+\textbf{C}),
\end{equation}
where $\textbf{C}=[C_{0},C_{1},\ldots,C_{N-1}]^T$ is the
peak-canceling signal vector in frequency domain. To avoid signal
distortion, the data vector $\textbf{X}$ and the peak reduction
vector $\textbf{C}$ lie in disjoint frequency domains, i.e.
\begin{equation}\label{eq5}
{X}_{n}+{C}_{n}=\left\{\begin{array}{lcr}{X}_{n}, &
n\in\mathcal{R}^C,\\ {C}_{n}, & n\in\mathcal{R},\end{array}\right.
\end{equation}
where $\mathcal{R}=\{i_{0},i_{1},\ldots, i_{M-1}\}$ is the index set
of the reserved tones, $\mathcal{R}^C$ is the complementary set of
$\mathcal{R}$ in $\mathcal{N} =\{0,1,\ldots,N-1\}$, and $M<N$ is the
size of PRT set.

The PAPR of the peak-reduced  OFDM signal
$\textbf{a}=[a_0,a_1,\ldots,a_{N-1}]^T$ is then
redefined~\cite{IEEEconf:18} as
\begin{equation}\label{eq6}
PAPR(\textbf{a})=\frac{\underset{0\leq n<
N}{\max}|x_{n}+c_{n}|^2}{E[|x_n|^2]}.
\end{equation}
Thus $\textbf{c}$ must be chosen to minimize the maximum of the
peak-reduced  OFDM signal $\textbf{a}$, i.e.
\begin{equation}\label{eq7}
\textbf{c}^*=\arg\underset{\textbf{c}}\min\underset{0\leq n<
N}\max|x_{n}+c_{n}|^2.
\end{equation}
To obtain the optimum $\textbf{c}^*$,  (\ref{eq7}) can be reformulated
as the following optimization problem:
\begin{equation}\label{eq8}
\begin{array}{rl}
\underset{\textbf{c}}\min & e,\\
\mbox{subject to} & e\geq 0,\\
& |x_n+c_n|^2\leq e,
\end{array}
\end{equation}
which is a Quadratically Constrained Quadratic Program (QCQP)
problem~\cite{IEEEconf:18} and $e$ is an optimization parameter.
 Although the optimum of a QCQP exists, the solution
requires a high computational cost. To reduce the complexity of the
QCQP,  a simple gradient algorithm proposed by Tellado
in~\cite{IEEEconf:16,IEEEconf:17,IEEEconf:18}  iteratively updates
the vector $\textbf{c}$ as follows:
\begin{equation}\label{eq9}
\textbf{c}^{(i+1)}=\textbf{c}^{(i)}-\alpha_i\textbf{p}[((k-k_i))_N],
\end{equation}
where $\alpha_i$ is a scaling factor,
$\textbf{p}=[p_0,p_1,\ldots,p_{N-1}]^T$ is called the time domain
kernel, and $\textbf{p}[((k-k_i))_N]$ denotes a circular shift of
$\textbf{p}$ to the right by a value $k_i$ calculated by
\begin{equation}\label{eq10}
k_i=\arg\underset{k}\max|x_k+c_k^{(i)}|.
\end{equation}
The time domain kernel $\textbf{p}$ is obtained by the following
formula:
\begin{equation}\label{eq11}
\textbf{p}=\textbf{Q}\textbf{P},
\end{equation}
where $\textbf{P}=[{P}_{0},{P}_{1},\ldots,{P}_{N-1}]^T$ is called
the frequency domain kernel whose elements are defined by
\begin{equation}\label{eq12}
{P}_{n}=\left\{\begin{array}{lcr}0, & n\in\mathcal{R}^C,\\ 1, &
n\in\mathcal{R},\end{array}\right.
\end{equation}
After $J$ iterations of this algorithm, the peak-reduced OFDM signal
is obtained:
\begin{equation}\label{eq13}
 \textbf{a}=\textbf{x}+\textbf{c}^{(J)}=\textbf{x}-\sum_{i=1}^{J}\alpha_i\textbf{p}[((k-k_i))_N].
\end{equation}

From (\ref{eq9})-(\ref{eq13}), it can be found that the PAPR
reduction performance of the TR-based OFDM system depends on the
selection of the time domain kernel $\textbf{p}$, which is only a
function of PRT set $\mathcal{R}$. When $\textbf{p}$ is a single
discrete pulse,  the best PAPR reduction performance can be obtained
because the maximal peak at location $k_i$ can be canceled without
distorting  other signal sampling. But it is impractical because a
single discrete pulse will results in that all tones should be
assigned to the PRT set. So we should select the time domain kernel
$\textbf{p}$ such that the $\textbf{p}$ not only reduces the peak at
location $k_i$ but also suppresses the other big values at location
$k\neq k_i$.

To find the optimal PRT set, in mathematical form, we require to
solve the following combinatorial optimization problem:
\begin{equation}\label{eq14}
\mathcal{R}^{*}=\arg\underset{\mathcal{R}}\min||[p_1, p_2,
\ldots,p_{N-1}]^T||_{\infty},
\end{equation}
which requires an exhaustive search of all combination of  possible
PRT set $\mathcal{R}$, i.e. $|\mathcal{R}|=C^{M}_{N}$ possible
combination numbers of  PRT set are searched, where
$C^{M}_{N}=\frac{N!}{M!(N-M)!}$ denotes the binomial coefficient. It
is an NP-hard problem  and cannot be solved for the  number of tones
envisaged in practical systems. In~\cite{IEEEconf:16,IEEEconf:18},
the consecutive PRT set, the equally spaced PRT set and the the
random PRT set optimization were proposed as the candidates of PRT
set. Although the consecutive PRT set and the equally spaced PRT set
are the simplest selections of PRT set, their PAPR reduction
performance are inferior to that of the random PRT set optimization.
But the random PRT set optimization requires larger PRT set
sampling, and the associated complexity limits the application of
such a technique. A variance minimization method
in~\cite{IEEEconf:19} is developed to solve the NP-hard problem, and
it is just a  modified version of the random PRT set optimization,
which also has the drawback of high computational cost.
In~\cite{IEEEconf:20}, a cross entropy method was proposed to solve
the problem. It obtains better results than the existing selection
methods, but it requires larger population or sampling sizes. These
limitations of the existing methods motivate us to find an efficient
method to obtain a nearly optimal PRT set. As mentioned before,  we
propose a genetic algorithm (GA) based PRT set selection method for
the purpose  with very low computational complexity.
\begin{figure} \centering
\includegraphics[width=3.5in,angle=0]{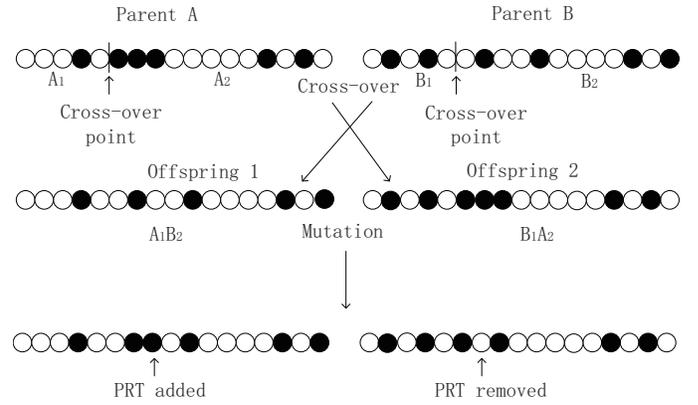}
\caption{An illustration of cross-over and mutation operations for $M=6$.}
\label{fig1}
\end{figure}

\section{Generic Algorithm Based PRT Set Selection}\label{sec:3}
In this section, we will briefly introduce the GA  and use it to search for a nearly optimal PRT set.
The resulting  PRT set will be used along with our proposed adaptive amplitude clipping technique
in the next section.

\subsection{A Brief Introduction to Genetic Algorithm}
The GA introduced by Holland~\cite{IEEEconf:24} is a
stochastic search method inspired from the principles of biological
evolution observed in nature. GA uses a population of candidate
solutions initially distributed over the entire solution space.
Based on the principle of  Darwinian survival of the fittest, GA
produces a better approximation to the optimal solution by evolving
this population of candidate solutions over successive iterations or
generations. The GA's evolution uses the following genetic
operators:
\begin{enumerate}
\item \emph{Selection} is a genetic operator that chooses a chromosome from
the current generation's population in order to include in the next
generation's mating pool. In general, chromosomes with a high
fitness (merit) should be selected and at the same time chromosomes
with a low fitness should be discarded.

\item \emph{Cross-over} is a genetic operator that exchanges the elements
between two different chromosomes (parents) to produce new
chromosomes (offsprings). The new population of the next generation
consists of these  offsprings.

\item \emph{Mutation} is a genetic operator that refers to the alteration of
the value of  each element  in a chromosome with a probability.
\end{enumerate}

GA has been applied to extensive optimization problems, such as
pilot location search  of OFDM timing synchronization
waveforms~\cite{IEEEconf:27}, joint multiuser symbol detection for
synchronous CDMA systems~\cite{IEEEconf:28}, the search of low
autocorrelated binary sequences~\cite{IEEEconf:29} and thinned
arrays~\cite{IEEEconf:30}. For a complete understanding of the GA,
the reader is referred
to~\cite{IEEEconf:24,IEEEconf:25,IEEEconf:26}.

\subsection{PRT Position  Search  Based on Genetic Algorithm}
A detailed description of the GA used for searching the nearly optimal
PRT set positions is described in what follows.

An initial population of $S$ chromosome (parent) sequences is
randomly generated. Each parent sequences is a vector of length $N$,
and each element of the  vector is a binary zero or one depending on
the existence of a PRT  at that position (one denotes existence and
zero denotes non-existence). The number of the PRT in each binary vector is
$M<N$. Denote the $S$ parent sequences as
$\{\textbf{A}_1,\dots,\textbf{A}_S\}$. Then each $\textbf{A}_\ell$
is a binary vector of length $N$.

For each parent sequence $\textbf{A}_\ell$, the PRT set $\cal
R_\ell$ is the collection of the locations whose elements are one.
Then the frequency domain kernel $\textbf{P}_\ell$ corresponding to
the PRT set $\cal R_\ell$ is obtained by (\ref{eq12}), and the time
domain kernel $\textbf{p}_\ell=[p_0^\ell,\dots,p_{N-1}^\ell]$ is
obtained by (\ref{eq11}). The merit (secondary peak) of the sequence
$\textbf{A}_\ell$ is defined as
\begin{equation}\label{merit}
m(\textbf{A}_\ell)=\parallel [p_1^\ell,\dots,p_{N-1}^\ell]^T
\parallel_\infty.
\end{equation}
%
The $T$ sequences (called elite sequences) with the lowest merits
are maintained for the next population generation. The best merit of
the $S$ sequences is defined as
\begin{equation}\label{bestmerit}
m^*=\min_{1\leq\ell\leq N}m(\textbf{A}_\ell).
\end{equation}

Then all $S$ sequences are crossed-over with a probability denoted
by $p_c$. For simplicity, one point cross-over is used in this
paper. In order to prevent local minima, mutation operator
controlled by a probability $p_m$ is applied by changing randomly
selected elements in a chromosome  sequence. Due to the cross-over
and mutation operations, the number of PRT in a newly generated
sequence (offspring) may be different to $M$ (the size of the PRT
set), so that the PRT set is infeasible. Hence one or more zero
(zeros) or one (ones) will replace the randomly selected elements
in the offspring to guarantee that the PRT set is feasible (the number
of the PRT in the offspring is $M$).

An illustration of the cross-over and mutation operations is
presented in Fig.~\ref{fig1}, where black circles represent PRT
positions. Chromosomes from two parents are separated from a
randomly selected point and crossed-over to generate new offsprings.
Due to crossed-over operation, the size of PRT set of an offspring
can be  more or less than the required $M$ PRTs. If an offspring has
PRTs more than the required $M$, then several  randomly selected
PRTs  will be removed. If an offspring has PRTs less than the
required $M$, then several PRTs  will be added to the randomly
selected positions.

The merits of all offspring sequences are evaluated using
(\ref{merit}).  Each sequence competes for the next generation pool.
The $T$ elite sequences obtained from the previous generation
replace the $T$ worst sequences with the highest merits in the
current generation. This increases the probability of generating
better solution and prevents the loss of the optimal solution
because of cross-over and mutation operations. The best merit of the
offsprings is evaluated using (\ref{bestmerit}), which will replace
the best merit of the previous generation if it is smaller than it.

The cycle is repeated until a predetermined number of times or a
solution  with a predefined fitness threshold (the merit is less
then some predefined threshold) is achieved. Therefore the proposed
GA-based PRT position search algorithm can be summarized in
Algorithm~1:

\begin{algorithm}
\caption{GA-PRT Algorithm}
\begin{algorithmic}
\STATE {1: Set the population size $S$, the PRT set size $M$, the
number $T$ of elite sequences, cross-over probability $p_c$,
mutation probability $p_m$ and the maximal iteration number $K$.}

\STATE {2: Randomly generate an initially feasible population of
size $S$, and find the PRT set $\cal R$ for each sequence. Calculate
the frequency domain kernel $\textbf{P}$ using (\ref{eq12}) and the
time domain kernel $\textbf{p}$ using (\ref{eq11}) for each
sequence.}

\STATE {3: Calculate the merits (secondary peaks) using
(\ref{merit}), select $T$ elite sequences with the lowest merits,
and find the best merit $m^*$ using (\ref{bestmerit}) and the
corresponding PRT set $\cal R^*$.}

\STATE {4: Cross-over and mutate all sequences by probabilities
$p_c$ and $p_m$ respectively, to generate a new feasible population
by randomly adding or removing PRTs.}

\STATE {5: Evaluate the merits (secondary peaks) and the best merit
of the new population. If the best merit of the new population is
smaller than that of the previous generation, then update the best
merit and the corresponding PRT set. Otherwise, remain the previous
best merit and the the corresponding PRT set  unchanged.}

\STATE {6: Replace the $T$ worst sequences with the highest merits
in the current generation by the $T$ elite sequences from the
previous generation and reselect $T$ elite sequences.}

\STATE {7: If the the maximal iteration number $K$ is achieved,
output the PRT set and the corresponding secondary peak, and
terminate the algorithm; Otherwise, go to Step 4.}

\end{algorithmic}
\end{algorithm}

\section{Adaptive Amplitude Clipping PAPR Reduction Algorithm}
Based on PRT set, some PAPR reduction methods have been developed.
The time domain gradient-based method proposed by Tellado
(TR-Gradient-Based Technique)~\cite
{IEEEconf:16,IEEEconf:17,IEEEconf:18} is of low complexity, but it
increases the signal average power  and requires very large
iterations to obtain the better solution.  Because the basic idea of
the gradient-based method comes from clipping,
TR-Clipping-Filtering-Based Technique is developed
in~\cite{IEEEconf:8}. This scheme iteratively clips the OFDM signal
to a predefined threshold $A$. The clipped signal is then  filtered
such that the clipping noise appears on the reserved tones only. But
the convergence speed of the method is  slow. An improved
adaptive-scaling TR (AS-TR) algorithm was proposed
in~\cite{IEEEconf:21, IEEEconf:22,IEEEconf:23}. The basic principle
of the AS-TR consists of two processes,  i.e. clipping in the time
domain and filtering in the frequency domain to suppress the peak
regrowth of the OFDM signal. Although the AS-TR provides a  better
PAPR reduction for predetermined clipping level,  the drawback of
the AS-TR is that the selection of the optimal clipping level is
very difficult. In practice, the optimal clipping level can not be
predetermined either. To overcome this  drawback of the AS-TR, we
propose a new TR-Clipping-Filtering-Based Technique for PAPR reduction.
Our method involves   adaptive amplitude clipping control, which allows  the
determination of the optimal clipping level. Simulation results
demonstrate that our scheme can obtain better PAPR reduction
regardless of the initial clipping level.

\subsection{A Brief Introduction to Adaptive-Scaling TR (AS-TR) Algorithm}
The  AS-TR method is an iterative clipping and filtered technique.
It firstly uses a soft limiter~\cite{IEEEconf:32} to the input OFDM signal to get the
clipping noise $f_{n}^{(i)}$,
\begin{equation}\label{eq15}
f_{n}^{(i)}=\left\{\begin{array}{lcr} {x_{n}^{(i)}-A}e^{j\theta_n^{(i)}}, & |x_{n}^{(i)}|>A,\\
0, & |x_{n}^{(i)}|\leq A,\end{array}\right.
\end{equation}
where $A$ is the target clipping threshold which is  relevant to the
clipping ratio   $\gamma=\frac{A^2}{E{|x_n|^2}}$, $\theta_{n}^{(i)}$
is the phase of $x_{n}^{(i)}$ and $i$ denotes the  iteration number.
The filtered clipping noise $\hat{f}_{n}^{(i)}$ is  obtained by
making $f_{n}^{(i)}$ through a filter whose passbands are only on
reserved tones. Let
$\textbf{f}^{(i)}=[f_{0}^{(i)},f_{1}^{(i)},\ldots,f_{LN-1}^{(i)}]^T$
and
$\hat{\textbf{f}}^{(i)}=[\hat{f}_{0}^{(i)},\hat{f}_{1}^{(i)},\ldots,\hat{f}_{LN-1}^{(i)}]^T$,
where $L$ is an oversampling factor.
The peak-reduction signal of the  AS-TR method
is iteratively updated as follows:
\begin{equation}\label{eq16}
\textbf{x}^{(i+1)}=\textbf{x}^{(i)}-\beta\hat{\textbf{f}}^{(i)},
\end{equation}
where $\beta$ is a positive  step size that determines the
convergence rate. The optimal $\beta$ is calculated by the following
formula:
\begin{equation}\label{eq17}
\beta^{(opt)}=\frac{\Re\left[\underset{n\in \textbf{S}_{p}}\sum
{f}_{n}^{(i)}\overline{\hat{f}_{n}^{(i)}}\right]}{\underset{n\in
\textbf{S}_{p}}\sum|\hat{f}_{n}^{(i)}|^2},
\end{equation}
where $\Re[x]$ represents the real part of $x$, $\textbf{S}_p=\{n:
n\in\textbf{S}_1$, $|x_n^{(i)}|>|x_{n-1}^{(i)}|$ and
$|x_n^{(i)}|\geq|x_{n+1}^{(i)}| \}$ is the index set of the peak of
$f_n^{(i)}$, and $\textbf{S}_1=\{n:|f_n^{(i)}|>0\}$ is the index set of
all clipping pulse.

In general, the larger PAPR reduction obtained by a lower target
clipping level is expected. But the PAPR reduction performance of
the AS-TR method is very sensitive to the target clipping level. In
other words, different clipping ratio $\gamma$  results in different
PAPR reduction performances, as will be demonstrated in Section V.
However, the optimal target  clipping level or clipping ratio can
not be predetermined at the initial stage. In the next section,
 an adaptive clipping scheme is proposed to get better PAPR reduction
 regardless of the initial  clipping ratio $\gamma$.

\subsection{Adaptive Amplitude Clipping   Algorithm for TR-Based OFDM Systems }
In this section, we propose an adaptive amplitude clipping algorithm
for TR-based OFDM systems. The main objective is to control both the
target clipping level $A$ and the convergence factor $\beta$ at each
iteration. The objective function is denoted as
\begin{equation}\label{eq18}
P=\underset{\beta,A}\min\underset{\textbf{S}_1}\sum
\left||x_{n}^{(i)}-Ae^{j\theta_n^{(i)}}|-\beta|\hat{f}_{n}^{(i)}|\right|^{2},
\end{equation}
where $\textbf{S}_1=\{n:|f_n^{(i)}|>0\}$ is the index of all
clipping pulses. The reason that we select (\ref{eq18}) as the
objective function is based on the following inequality.
\begin{equation}\label{eq19}
\left||x_{n}^{(i)}-Ae^{j\theta_n^{(i)}}|-\beta|\hat{f}_{n}^{(i)}|\right|\leq\left|x_{n}^{(i)}
-Ae^{j\theta_n^{(i)}}-\beta\hat{f}_{n}^{(i)}\right|.
\end{equation}
By least square method, (\ref{eq18}) shows that the optimal
convergence factor is
\begin{equation}\label{eq20}
\beta^{(i)}=\frac{\langle
|\textbf{f}^{(i)}|,|\hat{\textbf{f}}^{(i)}|\rangle}{\parallel
\hat{\textbf{f}}^{(i)}\parallel ^2},
\end{equation}
where $\langle\cdot,\cdot\rangle$ represents the real inner-product.
This implies that the calculation of $\beta$ involves real domain,
rather than  complex domain, which is another advantage of our
proposed algorithm.

Let $\textbf{S}_2=\{n:|f_n^{(i+1)}|>0\}$ and $\Omega=
\textbf{S}_1\bigcup\textbf{S}_2$. Suppose that the size of $\Omega$
is $N_1$. Then the gradient is updated as follows:
\begin{equation}\label{eq22}
\nabla_A=\frac{\underset{n\in\Omega}\sum
\left|f_n^{(i+1)}\right|}{N_1}.
\end{equation}
Then the proposed adaptive amplitude clipping (AAC-TR) algorithm is
stated in Algorithm~2.

\begin{algorithm}
\caption{AAC-TR Algorithm}
\begin{algorithmic}
\STATE {1: Set the initial  clipping level  $A$, the maximal
iteration number $K$, the step size $\rho$ and the reserved tone set
$\mathcal{R}$ obtained by Algorithm~1.}

\STATE{2: Set $i$=0, $\textbf{x}^{(0)}=\textbf{x}$ and $A^{(0)}=A$,
where
$\textbf{x}^{(i)}=[x_0^{(i)},x_1^{(i)},\ldots,x_{LN-1}^{(i)}]^T$.}

\STATE{3: Calculate the clipping noise $f_{n}^{(i)}$ using
(\ref{eq15}). If no clipping noise, transmit signal
$\textbf{x}^{(i)}$ and terminate the program.}

\STATE{4: Filter $f_{n}^{(i)}$ to satisfy the tone reservation
constraints:
\begin{enumerate}
\item [a)] Convert $\textbf{f}^{(i)}$ to
$\textbf{F}^{(i)}$=DFT$\{\textbf{f}^{(i)}\}$,

\item [b)] Obtain the
filtered clipping noise $\hat{\textbf{F}}^{(i)}$ by projecting
$\textbf{F}^{(i)}$ to the PRT set and remove the out-of band parts
of $\textbf{F}^{(i)}$,

\item [c)] Convert $\hat{\textbf{F}}^{(i)}$ to the time domain to obtain
$\hat{\textbf{f}}^{(i)}$ by carrying out the IDFT,
\end{enumerate}
}

\STATE{5: Calculate the optimal step size $\beta^{(i)}$  using
(\ref{eq20}), $\textbf{x}^{(i+1)}$ using (\ref{eq16})}, and
$\textbf{f}^{(i+1)}$ using (\ref{eq15}).

\STATE{6: Calculate $\nabla_A$ using (\ref{eq22}), and update the
clipping level $A$ by
\begin{equation}\label{eq21}
A^{(i+1)}=A^{(i)}+\rho\nabla_A.
\end{equation}
where $\rho$ is the step size with $0\leq\rho\leq1$.}
\STATE{7: Set $i=i+1$, if $i<K$, go to Step~3; Otherwise, transmit
$\textbf{x}^{(i+1)}$ and terminate the program.}
\end{algorithmic}
\end{algorithm}

\subsection{Complexity Analysis for AAC-TR Algorithm }
The complexity of the AAC-TR algorithm in oversampling case (oversampling factor $L\geq4$) for accurate PAPR
\cite{IEEEconf:10} is measured by using the number of real multiplications. A complex multiplication  is counted
as four real multiplications. We only consider the runtime complexity.  Step~1 and step~2
are not counted because all clipping-based methods must  carry out these two steps.

In step~3, $f_n^{(i)}$ can be calculated as $f_n^{(i)}=x_n^{(i)}-x_n^{(i)}(A/|x_n^{(i)}|)$, where $n\in\textbf{S}_1=\{n:|f_n^{(i)}|>0\}$,
and $N_{\textbf{S}_1}$ is the size of $\textbf{S}_1$.  Although $N_{\textbf{S}_1}$ is a random variable, it is roughly a constant
in all iterations and its mean can be calculated as follows~\cite{IEEEconf:21}.
 \begin{equation}\label{eq23}
\bar{N}_{\textbf{S}_1}=LNe^{-A^2/2{\sigma}^2}.
\end{equation}
So the complexity of computing $f_n^{(i)}$ can be estimated as
$2\bar{N}_{\textbf{S}_1}$ real multiplications, and
$\bar{N}_{\textbf{S}_1}$ real divisions.

In step~4, the number of real multiplications for computing an
$LN$-point DFT with $N_{\textbf{S}_1}$ nonzero inputs and $N$
in-band outputs (other outputs are not needed) can be computed as
follows~\cite{IEEEconf:23,IEEEconf:31}:
\begin{equation}\label{eq24}
\mathcal{M}_{LN}=\mathcal{M}_{LN/2}+2\mathcal{M}_{LN/4}\\+\max(0,\min(6N_{\textbf{S}_1},3LN/2-8))
\end{equation}
 \begin{equation}\label{eq25}
\mathcal{M}_{L}=0
\end{equation}
 \begin{equation}\label{eq26}
\mathcal{M}_{2L}=\max(0,\min(3N_{\textbf{S}_1},3L/2-4))
\end{equation}
where $\mathcal{M}_k$ represents the number of real multiplications
for computing a $k$-point DFT. Based on (\ref{eq24})-(\ref{eq26})
and replacing $N_{\textbf{S}_1}$ by $\bar{N}_{\textbf{S}_1}$, the
average complexity $\mathcal{M}_{DFT}$ of the DFT is calculated.
Similarly, replacing $N_{\textbf{S}_1}$, $N$ and $L$  by $M$, $LN$
and $1$ respectively, the average complexity $\mathcal{M}_{IDFT}$ of
the IDFT can be also calculated.

In (\ref{eq20}), the calculation of $\beta$ requires $2\bar{N}_{\textbf{S}_1}$ real multiplications and $1$
real division. Note that the update of $A$ in (\ref{eq21}) and the calculation of $\nabla_{A}$ in (\ref{eq22})
 only require $1$ real multiplication and $1$ real division, respectively.

The complexity of the AAC-TR algorithm mainly depends on the $LN$-point DFT/IDFT pair and weighting the
clipping noise in~(\ref{eq16}). The latter requires $2LN$ real multiplications. Based on the above analysis,
the total complexity of the the AAC-TR algorithm for $K$ iterations is
\begin{equation}\label{eq27}
\mathcal{M}=K(4\bar{N}_{\textbf{S}_1}+\mathcal{M}_{DFT}+\mathcal{M}_{IDFT}+2LN+1)
\end{equation}
real multiplications and $K(\bar{N}_{\textbf{S}_1}+2)$ real divisions.

If DFT/IDFT used in the AAC-TR algorithm is replaced by FFT/IFFT to
compute the peak-reduced signal, the computational complexity of the
AAC-TR algorithm is evaluated as $\mathcal{O}(LN\log (LN))$, which
is consistent with the AS-TR algorithm. But it is better than the
gradient algorithm~\cite{IEEEconf:16,IEEEconf:18}. The latter is
with complexity of order $\mathcal{O}(LN^2)$~\cite{IEEEconf:23}.
On the other hand, the AAC-TR algorithm can counteract all large
peaks above the clipping level in each iteration, while the gradient
algorithm  can only mitigate  one peak in each iteration.

Compared to the AS-TR algorithm, the complexity of the AAC-TR
algorithm  slightly increases due to the following factors. The
calculation of $\beta$ for the AS-TR algorithm in (\ref{eq17}) is
operated over $\textbf{S}_p$, which requires
$5\bar{N}_{\textbf{S}_p}$ real multiplications.  Nevertheless, the
calculation of $\beta$ for the AAC-TR algorithm in (\ref{eq20}) is
over $\textbf{S}_1$, which requires $2\bar{N}_{\textbf{S}_1}$ real
multiplications. From~\cite{IEEEconf:21}, we have
$\bar{N}_{\textbf{S}_1}=L\sqrt{\frac{6}{\pi}}\frac{\sigma}{A}\bar{N}_{\textbf{S}_p}$.
For example, when $L=4$, $N=512$, and $\gamma=5$\,dB, we
have $\bar{N}_{\textbf{S}_1}=86.6902 $ and
$\bar{N}_{\textbf{S}_p}=39.4389 $.
So $5\bar{N}_{\textbf{S}_p}-2\bar{N}_{\textbf{S}_p}=23.8139$ real multiplications
are reduced in each iteration. Although the adaptive update of clipping level in
(\ref{eq21}) and (\ref{eq22}) will result in the increment of
calculation (mainly real additions), the operation reduction of
computing $\beta$ can compensate some of such increment so that the
complexity of the AAC-TR algorithm  slightly increases compared to
that of the AS-TR algorithm.

\section{Simulation results}
To evaluate and compare the performance of GA based nearly optimal PRT
set positions searching and the AAC-TR algorithm for OFDM PAPR
reduction, extensive simulations have been conducted. In our
simulations, an OFDM system of 16-QAM (quadrature amplitude modulation)  with $N=512$ sub-carriers is used.
The number of reserved PRT set is $M=32$.
 In order to get CCDF, $10^5$ random OFDM symbols are generated. The
transmitted signal is oversampled by a factor of $L=4$ for accurate PAPR estimation.

In the GA-PRT algorithm, the population size $S=30$, the maximum
iteration number $K=170$, the cross-over probability $p_c=0.9$, the
mutation probability $p_m=0.05$, and the elite sequences $T=2$. For
comparison, the random set optimization (RSO) and CE algorithm are
also tested. The  optimal PRT set of  the RSO is obtained by
generating $10^5$ random sets and selecting the best PRT set.  The
parameters used in the CE algorithm basically
follow~\cite{IEEEconf:20}, i.e. the population size or the number of
samples is $U=120$, the step size $\rho=0.1$, the smoothing factor
$\lambda=0.8$, and the  maximum iteration number $K=170$.

The corresponding PRT sets obtained by the proposed GA-PRT algorithm and the
existing methods for $M=32$ are as follows.
\begin{multline}
{\mbox{GA-PRT}}=\{10,11,28,42,43,61,95,107,115,120,131,155,\\~~~~~~~~~~~~~~~~~~~~156,160,176,193,202,215,
232, 254,273,316,321,\\337,370,384, 403,412,416,447,484,485\},
\end{multline}
\begin{multline}
{\mbox{CE-PRT}}=\{8,49,75,76,111,117,127,134,145,156,159,\\~~~~~~~~~~~~~~~~~~~~163,164,202,214,223,
258,268,273, 322,342,350,\\~~~~~~~~~~~~~~~~~~~~366,412,427,438,455,457,458,467,488,504\},
\end{multline}
\begin{multline}
{\mbox{CS-PRT}}=\{225,226,227,228,229,230,231,232,233,234,\\~~~~~~~~~~~~~~~~~~~~~235,236,237,238,239,
240,241,242, 243,244,245,\\~~~~~~~~~~~~~~~~~~~~~246,247,248,249,250,251,252,253,
254,255,256\},
\end{multline}
\begin{multline}
{\mbox{ES-PRT}}=\{15,31,47,63,79,95,111,127,143,159,175,191,\\~~~~~~~~~~~~~~~~~~~~~207,223,239,255,271,
287,303,319,335,351,367,\\~~~383,399,415,431,447,463,479,495,511\},
\end{multline}
\begin{multline}
{\mbox{RS-PRT}}=\{16,45,57,61,63,80,81,104,105,118,134,155,\\~~~~~~~~~~~~~~~~~~~~~159,167,184,187,198,
200,201,203,241,250,276,\\~~~~284,329,394,408,459,466,481,495,498\},
\end{multline}
where GA-PRT, CE-PRT, CS-PRT, ES-PRT and RS-PRT respectively
represent GA optimization based PRT set, CE optimization based PRT
set, consecutive PRT set, equally spaced PRT set and random set
optimization based PRT set.

\begin{figure}
\centering
\includegraphics[width=3.5in,angle=0]{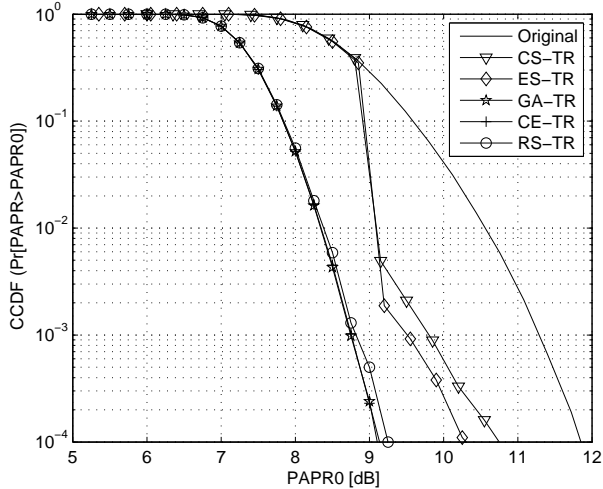}
\caption{Comparison of PAPR reduction based on the Tellado's gradient algorithm with different PRT set.}
\label{fig2}
\end{figure}

\begin{table}[h]
\begin{center}
\caption {Comparison of computational complexity (CC) and secondary peak (SP)
for different methods, and the difference of the secondary peaks obtained
by CE method and other methods}\label{Table1}
\begin{tabular}{|c|c|c|c|}

\hline    methods    & CC                          & SP           &differences\\
\hline     CS-PRT      & -                          &0.9936        & 0.7131\\
\hline     ES-PRT      &-                           &1              &0.7195\\
 \hline    GA-PRT      &$SK=30*170=5100$           &0.2996         &0.0191\\
\hline     CE-PRT      & $UK=120*170=20400$        &0.2805         &0\\
\hline     RS-PRT      &  $10^5$                   &0.3207         &0.0402\\
 \hline
\end{tabular}
\end{center}
\end{table}
\subsection{Computational Complexity Versus Normalized Secondary Peak}
Table~\ref{Table1} compares the  computational complexity
and the normalized secondary peak among different methods. For
comparison, the differences between the normalized secondary peak of the
CE algorithm and those of other methods are also calculated. It can
be seen that the secondary peaks achieved by consecutive PRT set and
equally spaced PRT set are the worst among all methods. The
secondary peak obtained by the CE algorithm excels those of other
methods. The secondary peak by random set optimization (RSO) with
$10^5$ randomly chosen PRT set is $0.0402$ larger than that of CE
algorithm with $20400$ searches. So the CE algorithm with lower
computational complexity gets better performance than RSO. On the
other hand,  the computational complexity of CE
algorithm is five times greater than that of GA-PRT algorithm. But the
difference between the secondary peaks of the two methods is only
$0.0191$. In other words, the complexity reduction by GA-PRT algorithm
is $(20400-5100)=15300$ operations with payment of only $0.0191$ in
the secondary peak. So the proposed GA-based PRT set selection
algorithm is more efficient than the  CE algorithm for solving the
secondary peak. The fact that the PAPR reduction performances of CE
and GA are almost the same will be demonstrated in Fig.~\ref{fig2}.

\subsection{PAPR Reduction Versus Different PRT Sets}
In Fig.~\ref{fig2}, the comparison of PAPR reduction performance
based on Tellado's gradient algorithm
(GD-TR)~\cite{IEEEconf:16,IEEEconf:18} with the above different PRT
sets is shown. Here the iteration number of the gradient algorithm
is set to $10$.  When $P_r(PAPR>PAPR_0)=10^{-4}$, the PAPR of the
original OFDM is $12$\,dB. The PAPRs of  CS-PRT set and ES-PRT set
are $10.8$\,dB and $10.5$\,dB, respectively. Using the random set
optimization in~\cite{IEEEconf:18}, when the number of randomly
selected PRT sets is $10^5$, the PAPR is reduced to $9.2$\,dB. The
PAPR obtained by the CE-PRT with the search complexity
$UK=120*170=20400$ in~\cite{IEEEconf:20} is $9.1$\,dB. The PAPR
obtained by the GA-PRT with the search complexity $SK=30*170=5100$
is $9.1$\,dB. There is a negligible  gap between the PAPRs obtained
by CE-PRT and by GA-PRT. But from Table~\ref{Table1}, we see that
the search complexity of the GA-PRT is only $SK/UK=30/120=25\%$ of
that by the CE-PRT.

\begin{figure}
\centering
\includegraphics[width=3.5in,angle=0]{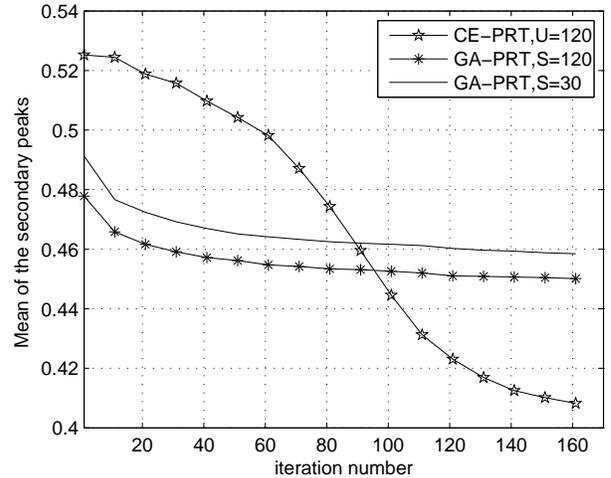}
\caption{Comparison of the mean of the best secondary peak between the GA-PRT algorithm and the CE-PRT one.} \label{fig3}
\end{figure}

\subsection{Comparison of the Secondary Peak Between GA-PRT Algorithm and CE-PRT Algorithm}
In Fig.~\ref{fig3}, 100 experiments are performed to compare  the
means of the best secondary peak obtained by GA-PRT algorithm and
CE-PRT algorithm. According to the original CE algorithm proposed by
Rubinstein, the sample size $U$ is very large to get better
performance for  the CE-PRT algorithm, so we only adopt $U=120$ used
in the~\cite{IEEEconf:20} for comparison. It is can be found that
the  performance of the GA-PRT algorithm is better than that of the
CE-PRT one in approximately 1-90 iterations. As the increase of
iterations, the secondary peaks of the CE-PRT algorithm are
improved. This displays that the convergence of the CE-PRT algorithm
is slower than that of the GA-PRT one, so that the proposed GA-PRT
algorithm can get a better suboptimal PRT set. On the other hand,
the maximal difference of the secondary peak gained by the GA-PRT
algorithm between $S=30$ and $S=120$ is only 0.0134, so $S=30$ is a
better choice for the proposed GA-PRT algorithm.

\begin{figure}
\centering
\includegraphics[width=3.5in,angle=0]{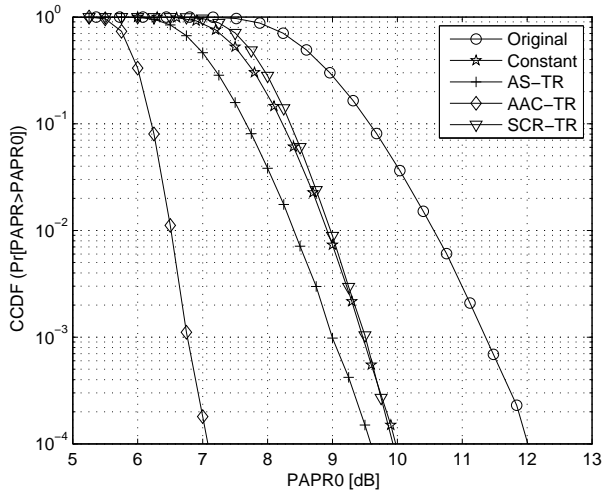}
\caption{Comparison of PAPR reduction for different methods with the same GA-PRT set.} \label{fig4}
\end{figure}

\begin{figure} \centering
\includegraphics[width=3.5in,angle=0]{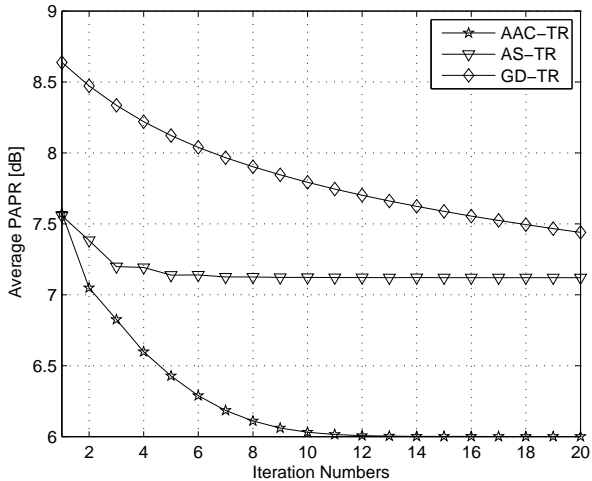}
\caption{Relationship of PAPR reduction with iteration numbers for different methods.}
\label{fig5}
\end{figure}
\subsection{PAPR Reduction Versus Different Methods}
 Fig.~\ref{fig4} compares the PAPR reduction performance of the
proposed AAC-TR algorithm with the  constant scaling (Constant)
algorithm, AS-TR method in~\cite{IEEEconf:21, IEEEconf:22}, signal
to clipping noise ratio (SCR-TR) algorithm and gradient descent
(GD-TR) algorithm~\cite{IEEEconf:17} for the same GA-PRT set. Here
the iteration number of the  constant scaling  algorithm is  $40$.
The same maximum iteration number is set to $10$ for AS-TR, AAC-TR,
GD-TR and SCR-TR. When $P_r(PAPR>PAPR_0)=10^{-4}$, the PAPR of the
original OFDM is $11.9$\,dB. Using the constant scaling  algorithm
to SCR-TR algorithm GD-TR algorithm and AS-TR algorithm, the PAPRs
are approximately reduced to
  $9.33$\,dB, $9.67$\,dB, $9.22$\,dB, $8.56$\,dB, respectively.
 The PAPR is approximately reduced to $7.05$\,dB by using the proposed
AAC-TR algorithm. Compared to the PAPR of the original OFDM, an
approximate $4.85$\,dB reduction gain is obtained, which is $1.51$\,dB
larger than AS-TR algorithm,  $2.62$\,dB larger than SCR-TR
algorithm, $2.17$\,dB larger than GD-TR algorithm and $2.28$\,dB
larger than constant scaling algorithm for the same $10$
iterations.

\subsection{Average PAPR Reduction Versus Iteration}
 Fig.~\ref{fig5} compares the average PAPR reduction
performance of the AS-TR, AAC-TR and GD-TR with clipping ratio
$\gamma=4$\,dB for the same iteration numbers.  Fig.~\ref{fig4}
shows that the average PAPR reduction performance of
the AAC-TR algorithm is better than those of AS-TR and GD-TR. When
the iteration number equals $11$, the AAC-TR algorithm converges to
$6$\,dB PAPR.  Although the GD-TR algorithm  is simple, its
convergence speed is the slowest among the three methods. When the
iteration number is $20$, its average PAPR is approximately
$7.4$\,dB, which is  $1.4$\,dB larger than AAC-TR algorithm in $11$
iterations. The AS-TR algorithm converges to $7.1$\,dB PAPR in $7$
iterations, however, which is approximately $0.9$\,dB larger than
AAC-TR algorithm in the same iterations.

\begin{table}[h]
\begin{center}
\caption {Comparison of average power increase (API), average simulation time (AST) and
PAPR  for different methods with $\gamma=5$\,dB}\label{Table2}
\begin{tabular}{|c|c|c|c|}

\hline    methods    & API (dB)     &  AST (ms) &PAPR (dB)\\
\hline     GD-TR      &0.3867        & 7.7485   & 9.22  \\
 \hline    AS-TR      &0.3854        & 11.3914   &8.56\\
\hline     AAC-TR     &0.2596        & 13.2898   &7.05\\
 \hline
\end{tabular}
\end{center}
\end{table}


\subsection{Average Power Increase, Average Simulation Time and PAPR Reduction Versus Different Methods}
Table~\ref{Table2} compares the  average power increase (API)
(in dB), average simulation time (AST) (in millisecond) and PAPR for
AS-TR, AAC-TR and GD-TR with clipping ratio $\gamma=5$\,dB for $10$
iterations. We observe  that the AST of the GD-TR method is least
(Note that the GD-TR method must prestore an $LN\times LN$ IFFT matrix,
the calculation does not include in simulation time),
but its PAPR is $0.66$\,dB larger than that of the AS-TR algorithm,
and $2.17$\,dB larger than that of the AAC-TR algorithm. The APIs of
 AS-TR and AAC-TR  are almost the  same. Compared to AS-TR algorithm, the AST
of the AAC-TR algorithm increases slightly, but its API is less than that
of the AS-TR, and PAPR is $1.51$\,dB smaller than
that of AS-TR.

\subsection{ PAPR Reduction Versus Different Clipping Ratios}
\begin{figure} \centering
\includegraphics[width=3.5in,angle=0]{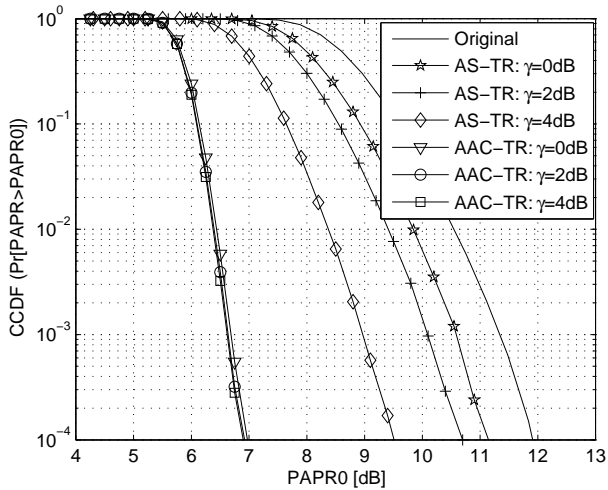}
\caption{Comparison of PAPR reduction between  AAC-TR and AS-TR with
different clipping ratio.} \label{fig6}
\end{figure}

Fig.~\ref{fig6} compares the PAPR reduction performance of AS-TR
algorithm and AAC-TR method with $10$ iterations for three different
target clipping ratios, $\gamma=0$\,dB, $2$\,dB and $4$\,dB.  For
comparison, the original OFDM signal's PAPR is also given.  When
$P_r(PAPR>PAPR_0)=10^{-4}$, the AS-TR algorithm  for the three
different clipping ratios, $\gamma=0$\,dB, $2$\,dB and $4$\,dB can
obtain $0.9$\,dB, $1.4$\,dB and $2.4$\,dB PAPR reduction from the
original OFDM PAPR of $12$\,dB. It demonstrates that the AS-TR
algorithm is  sensitive to  the target clipping ratio. Different
target clipping ratios can result in different PAPR reduction
performance. Contrary to the AS-TR algorithm, the proposed AAC-TR
algorithm obtains an approximately $5$\,dB PAPR reduction from the
original OFDM PAPR of $12$\,dB for all three of the target clipping
ratios.

\subsection{Different $\rho$ Versus PAPR Reduction}
\begin{figure} \centering
\includegraphics[width=3.5in,angle=0]{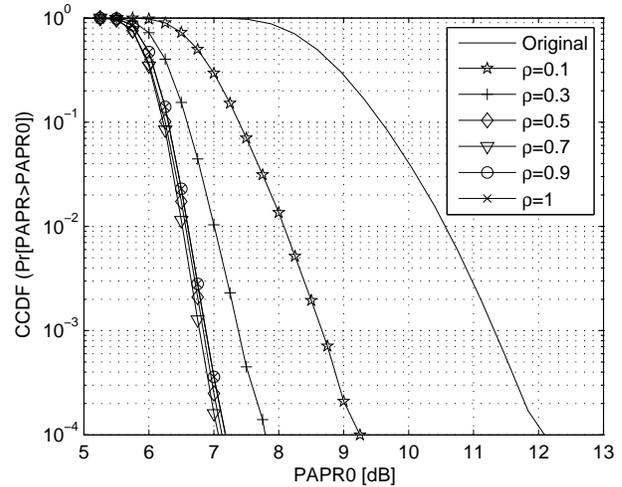}
\caption{Comparison of PAPR reduction with different step size $\rho$.}
\label{fig7}
\end{figure}

In Fig.~\ref{fig7}, we compare the PAPR reduction performance of AAC-TR
algorithm for different step size $\rho$. When $\rho=0.1$ and $\rho=0.3$,
the PAPRs are $9.3$\,dB and $7.8$\,dB. For other choices on $\rho$,
the differences of the PAPRs are very small. The reasons  can be that
the smaller $\rho$ can not effectively adjust the clipping level $A$. This corresponds
to that $A$ is not updated. So we should select a bigger  step size $\rho$ to gain better
PAPR performance for the AAC-TR algorithm.

\section{Conclusion}
This paper studies  PAPR reduction for tone reservation-based OFDM
systems.  The PAPR reduction  mainly depends on the selection of
peak reduction tone (PRT) set and the optimal target clipping level.
Finding the optimal PRT set is equivalent to solving  the secondary
peak minimization problem, which must optimize over  all combination
of  possible PRT sets. It is an NP-hard problem   and cannot be
readily solved for the  number of tones appeared in practical OFDM
systems.  The existing selection methods, such as the consecutive
PRT set, the equally spaced PRT set and the random PRT set, perform
poorly compared to  the optimal PRT set or require high
computational complexity. In this paper, an efficient scheme based
on genetic algorithm (GA) was proposed to give a  nearly optimal PRT
set. Compared to the CE-PRT algorithm, the proposed GA-PRT algorithm
has lower computational complexity and achieves a good approximation
to the secondary peak of the CE-PRT algorithm. Although the
TR-clipping-based technique is simple and attractive for practical
implementation, finding the optimal target clipping level is
difficult and the optimal target clipping level can not be
predetermined in the initial stage. An adaptive clipping control
algorithm is  proposed to solve this problem. Simulation results
show that the proposed  adaptive clipping control algorithm achieves
good  PAPR reductions regardless of the target clipping ratios.

\end{document}